\documentclass[a4paper, 12pt, english]{article}

\usepackage[utf8]{inputenc}
\usepackage{amsmath,amssymb}
\usepackage{graphicx}
\usepackage{subfig}
\usepackage[colorinlistoftodos]{todonotes}

\usepackage{amsmath}
\usepackage{amssymb}
\usepackage{graphicx}
\usepackage{algorithm2e}
\usepackage{breqn}
\usepackage{multirow}
\usepackage{capt-of}
\usepackage[T1]{fontenc}
\usepackage{letltxmacro}
\usepackage{listings}
\usepackage{color}

\definecolor{dkgreen}{rgb}{0,0.6,0}
\definecolor{gray}{rgb}{0.5,0.5,0.5}
\definecolor{mauve}{rgb}{0.58,0,0.82}

\usepackage{indentfirst}
\usepackage{verbatim}
\usepackage{textcomp}
\usepackage{gensymb}

\usepackage{relsize}

\usepackage{lipsum}
\usepackage{xcolor}
\usepackage{xparse}
\NewDocumentCommand{\myrule}{O{1pt} O{2pt} O{black}}{%
  \par\nobreak 
  \kern\the\prevdepth 
  \kern#2 
  {\color{#3}\hrule height #1 width\hsize} 
  \kern#2 
  \nointerlineskip 
}
\usepackage[section]{placeins}

\usepackage{booktabs}
\usepackage{colortbl}%

\usepackage[obeyspaces]{url}
\usepackage{etoolbox}
\usepackage[colorlinks,citecolor=black,urlcolor=blue,bookmarks=false,hypertexnames=true]{hyperref} 

\usepackage{geometry}
\begin{document}

\begin{titlepage}
\begin{center}

\includegraphics[width=0.5\textwidth]{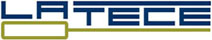}\\
\textbf{\large Laboratoire de Recherches sur les Technologies du Commerce Électronique}\\[0.2cm]

\vspace{10pt}

\includegraphics[width=0.5\textwidth]{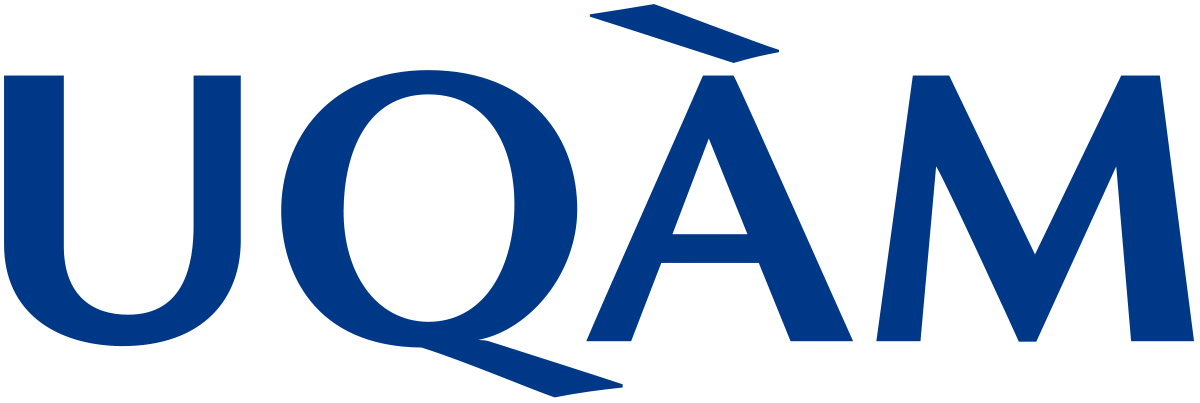}\\[1cm]

\textbf{\LARGE Universit\'e du Qu\'ebec \`a Montr\'eal}\\[0.5cm] 
\vspace{20pt}

\par
\myrule[1pt][7pt]
\textbf{\Large A Static Program Slicing Approach for Output Stream Objects in JEE Applications}\\
\myrule[1pt][7pt]

\vspace{35pt}
\large{\textbf{Anas Shatnawi, Hafedh Mili, Manel Abdellatif, \\Jean Privat, Yann-Ga\"el Gu\'eh\'eneuc, Naouel Moha, Ghizlane El Boussaidi}}\\[0.3cm]

\textit{LATECE Technical Report  2017-5, LATECE Laboratoire, Universit\'e du Qu\'ebec \`a Montr\'eal, Canada}

\vspace{150pt}
\small{July, 2017}

\end{center}

\par
\vfill
\begin{center}
\end{center}
\end{titlepage}



\newpage

\begin{center}
\textbf{\Large A Static Program Slicing Approach for Output Stream Objects in JEE Applications}\\[0.9cm]

{Anas Shatnawi\footnote{anasshatnawi@gmail.com}, Hafedh Mili\footnote{mili.hafedh@uqam.ca}, Manel Abdellatif, Jean Privat, \\Yann-Ga\"el Gu\'eh\'eneuc, Naouel Moha, Ghizlane El Boussaidi}\\[0.5cm]

\textit{LATECE Technical Report 2017-5, LATECE Laboratoire, Universit\'e du Qu\'ebec \`a Montr\'eal, Canada}

\end{center}

\begin{abstract}
	In this paper, we propose a program slicing approach for the output stream object in JEE applications. Our approach is based on extracting a dependency call graph from KDM models of JEE applications. Then, it applies breath-first search algorithm to identify the program slice as a graph reachability problem. The proposed approach is implemented as an extension of our DeJEE tool.
\end{abstract}

\section{Introduction}
\label{Introduction}
\subsection{Program Slicing}
Program slicing techniques support several software engineering tasks including, but are not limited to, understanding, maintenance, evolution, change impact analysis and reengineering \cite{weiser1981program}.

\begin{figure}[h]
	\begin{center}
		\includegraphics[width=0.60\textwidth]{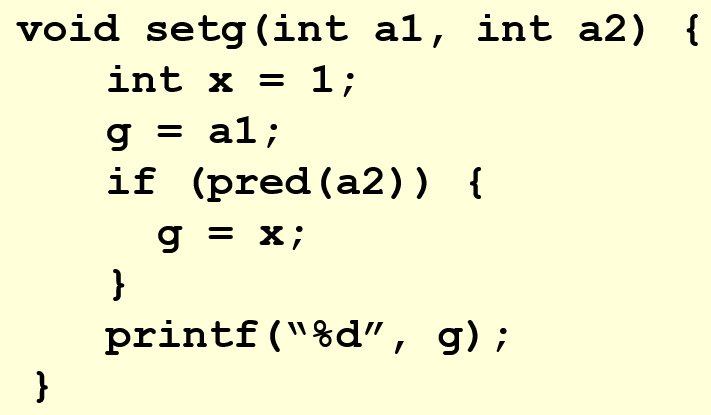}
		\caption{An example of a program}
		\label{fig:program}
	\end{center}
\end{figure}

\begin{figure}[h]
	\begin{center}
		\includegraphics[width=0.60\textwidth]{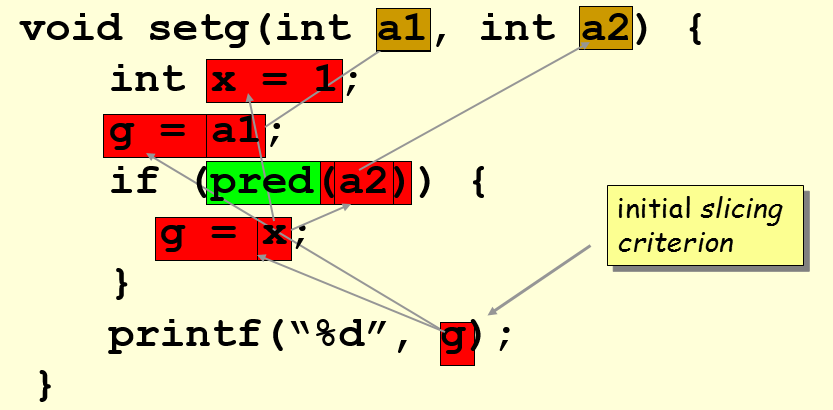}
		\caption{A program slice of the \textit{g} variable}
		\label{fig:program-slice}
	\end{center}
\end{figure}

The concept of program slicing was firstly introduced by Mark Weiser in 1981 \cite{weiser1981program}.
Following Weiser's definition \cite{weiser1981program}, a \textit{program slice} is defined in terms of a group of statements that impact the value of a given program variable at a point of interest, during the program execution. Considering the example in Figure \ref{fig:program}, the program slice of the \textit{g} variable is presented in Figure \ref{fig:program-slice}.
The identification of a program slice can be based on either static \cite{nguyen2015cross} or dynamic \cite{zhang2003precise} analysis techniques.

\subsection{Problem with Program Slicing in JEE Applications}
Identifying a program slice is a challenge in JEE applications that combine server-side Java code with a number of client-side Web dialects (e.g., HTML, JSP, JSF), with a number of dependencies within and between different languages embodied in container services, various ad-hoc configuration files and string literals.

\begin{figure}[h]
	\begin{center}
		\includegraphics[width=\textwidth]{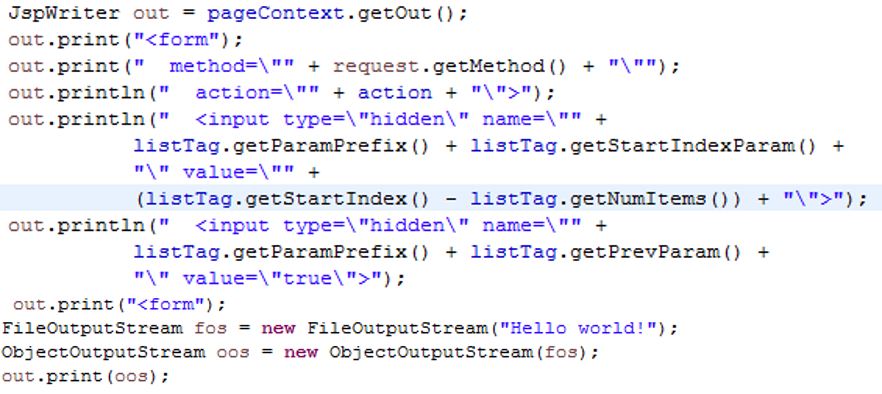}
		\caption{Example of mixing server-side with client-side code in the output stream object}
		\label{fig:output-stream-example}
	\end{center}
\end{figure}

A common example in JEE applications is Java \textit{Servlets} and \textit{Tag Handlers} that use a special \textit{output} stream object, offered by the Web container, to send data to their clients. Figure \ref{fig:output-stream-example} shows an example taken from a tag handler.
Identifying a program slice of such output stream objects is not a trivial due to several challenges. Such challenges are: 

\begin{itemize}
\item The combination of string literals, HTML tags, object-oriented method invocations and attribute accesses as parameters attached to these objects.

\item The Java standard allows output stream objects to be built based on nested stream and writer instances of \textit{BufferedStream}, \textit{PrintWriter}, \textit{ObjectOutputStream}, etc.

\item The may use of Java reflexion.

\item The may use of virtual method dynamic dispatching (e.g., polymorphism).

\item They may rely on user input data.
\end{itemize}

\subsection{Proposition}
In this paper, we propose a static program slicing approach\footnote{In our approach, we focus on solving the first two problems mentioned previously} to identify a minimal program slice of the output stream object of a given Servelt/Tag Handler. Our approach aims to extract the set of statements that impact the value of this output stream object based on the analysis of KDM models. The use of KDM models allows us to generalize our approach to be for multilanguage applications. The proposed approach is implemented as an extension of our DeJEE tool \cite{shatnawi2017analyzing}.

\subsection{Paper Organization}
The rest of this paper is organized as follows. We present an overview of the proposed approach in Section \ref{sec:overview}. Then, we discuss our approach in Section \ref{sec:build-depndency-call-graph} and Section \ref{sec:identify-program-slcie} that aims to identify a dependency call graph and to identify a program slice from this dependency call graph respectively. In Section \ref{sec:tool-implementation}, we talk about the implementation of our approach. Last, we conclude this paper in Section \ref{sec:conclusion}.

\section{Overview of the Proposed Slicer}
\label{sec:overview}
Our program slicing approach aims to identify the set of program statements that contribute in the output stream object of a given Servlet/Tag Handler. Such statements are:
\begin{enumerate}
\item Method invocations that write parameters in the output stream object. We identify three methods: 

\begin{enumerate}
\item The \textit{Print} method, e.g., \textit{out.print(parameter);}, with respect to the other versions of this print method. 
\item The \textit{Write} method, e.g., \textit{out.write(parameter);}
\item The \textit{Append} method, e.g.,  \textit{out.append(parameter);}
\end{enumerate}

\item The set of statements that have either direct or indirect impact of the parameters of the writing methods of the object. This includes: (a) method invocations, (b) variables, (c) control statements. 
\end{enumerate}

To identify these statements, we develop a process of two steps:

\paragraph{Step 1: Build Dependency Call Graph} We want to identify relationships between the program statements of a given method/class. To this end, we build a dependency call graph between statements.

\paragraph{Step 2: Identify Program Slice of Output Stream Object} We use this dependency call graph to identify the program slice of the output stream object. We rely on a graph reachability algorithm to identify the set of nodes (statements) that are related to the value of the output stream object. These statements consist of all statements that the output stream object relies on either directly or indirectly.

\section{Build Dependency Call Graph}
\label{sec:build-depndency-call-graph}
\subsection{The Definition of Dependency Call Graph}
Our dependency call graph is a \textit{directed graph G = $\langle V, E\rangle$}, where \textit{V} is the set of nodes corresponding to program statements and \textit{E} is the set of edges between these nodes/statements \textit{V}, such that an edge links the nodes/statements \textit{(u,v)} if \textit{u} depends on \textit{v}. \textit{u} depends on \textit{v} based on at least one of these two types of dependencies: data and control dependencies.

\subsubsection{Data dependencies} 

It refers to the read/write relationships between statements. All statements that read/reference a given variable(s) depend on all statements that write/change the value of this variable(s) with respect to the condition that the execution of write statements should be before the execution of read ones. 

Thus, we consider that a statement \textit{u} depends on a statement \textit{v} if \textit{u} read the value of at least one variable that has been previously modified by \textit{v}. For example, considering  that the \textit{x=y;} statement is executed before the \textit{z=x;} one. Then, \textit{z=x;} depends on \textit{x=y;}.

\subsubsection{Control dependencies} 

It refers to the relationships that are existed between the conditional statements and their inner statements. Conditional statements control whether their inner statements will be executed or not. 

Thus, we consider that \textit{u} depends on \textit{v} if \textit{u} is an inner statement of the conditional statement \textit{v}. 

\begin{figure}[h]
	\begin{center}
		\includegraphics[width=\textwidth]{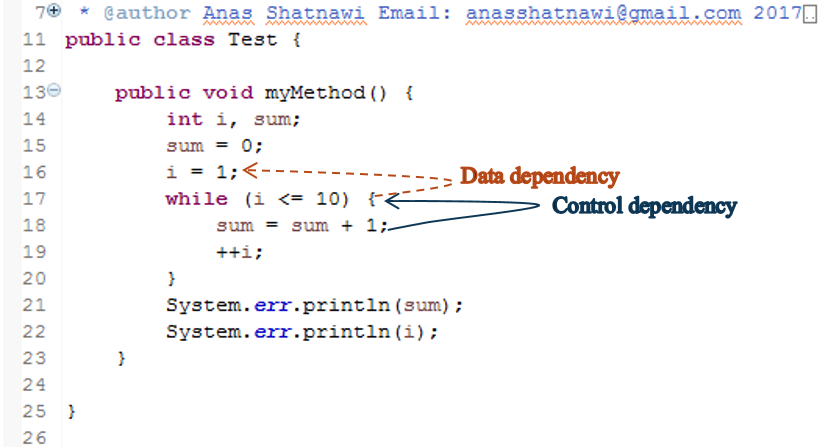}
		\caption{Example of data and control dependencies}
		\label{fig:dependency-example}
	\end{center}
\end{figure}

Figure \ref{fig:dependency-example} shows an example of these two kinds of dependencies.

\subsection{The Procedure of Building Dependency Call Graph}
We propose a procedure for building a dependency call graph of a given method. This procedure is as follows:

\begin{enumerate}
\item We create a node for each statement in this method including the method's prototype. For our example in Figure \ref{fig:dependency-example}, each line will be considered as a node.

\item We add the control dependencies as follows: 

	\begin{enumerate}
		\item We consider the node corresponding to the method's prototype as an entry point for the graph. 		Thus, we create control dependencies from this node going to all other nodes. 
        
        For example, in Figure \ref{fig:dependency-example}, the node of \textit{myMethod()} will have direct links to all nodes.
        
        \item For each control statement (e.g., \textit{for}, \textit{while}, \textit{if}, \textit{switch}, etc.), we create control dependencies from each node of a control statement going to all nodes corresponding to its inner statements. 
        
        For example, links are added from the node of \textit{while (i <= 10)} to the nodes of \textit{sum = sum + 1;} and \textit{++i;}, in Figure \ref{fig:dependency-example}. 

\end{enumerate}

\item We add data dependencies by computing read/write relationships between statements. 
We create a dependency from the node that its statement writes/changes the value of a variable to all nodes corresponding to statements that read the value of this variable. We detect the value change of a variable based on the assignment operator (i.e., \textit{=}). 

To make sure that write statements are executed before the write ones, we gradually evaluate statements following their sequential positions in the program.
Each time we evaluate a node (statement), we check the variable(s) that it reads with the variables that have been written in the already evaluated statements.
\end{enumerate}

\begin{figure}[h]
	\begin{center}
		\includegraphics[width=\textwidth]{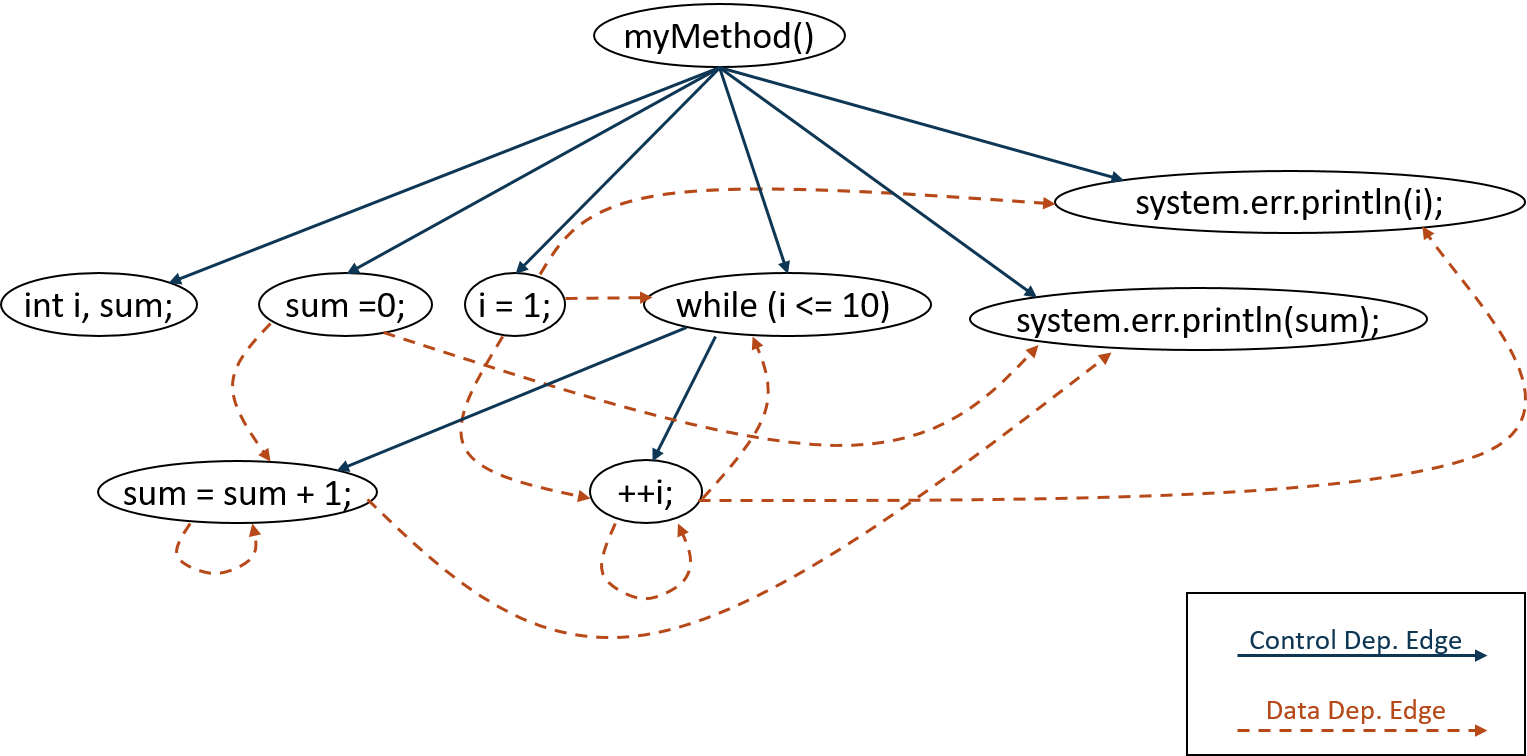}
		\caption{Example of a dependency call graph of the source code in Figure \ref{fig:dependency-example}}
		\label{fig:dependency-call-graph example}
	\end{center}
\end{figure}

Figure \ref{fig:dependency-call-graph example} shows the dependency call graph resulting from the source code presented in Figure \ref{fig:dependency-example}.

\section{Identify Program Slice of Output Stream Object}
\label{sec:identify-program-slcie}
In the previous section, we identified a dependency call graph that shows the program dependencies between statements based on their control and data flow. To identify the program slice by formulating the problem as a graph reachability problem, we need to identify a \textit{transpose(reverse) graph} of the identified dependency call graph. Then, we apply a \textit{Breadth-First Search} algorithm to identify the set of node that are reachable from a given node.
In the remaining sub-sections, we explain these two steps.

\subsection{Identifying Transpose Graph of the Dependency Call Graph}
We identify the \textit{transpose}\footnote{\textit{G$^{T}$} is computable in \textit{O(|V| + |E|)} time } graph \textit{G$^{T}$ = (V, E$^{T}$)} of our dependency call graph \textit{G = (V, E)}. \textit{G$^{T}$} contains the same set of nodes \textit{V} of \textit{G}, but it reverses the direction of each edge in \textit{G} such that \textit{E$^{T}$ = \{(u, v) | (v, u) $\in$ E\}}. 

\begin{figure}[h]
	\begin{center}
		\includegraphics[width=\textwidth]{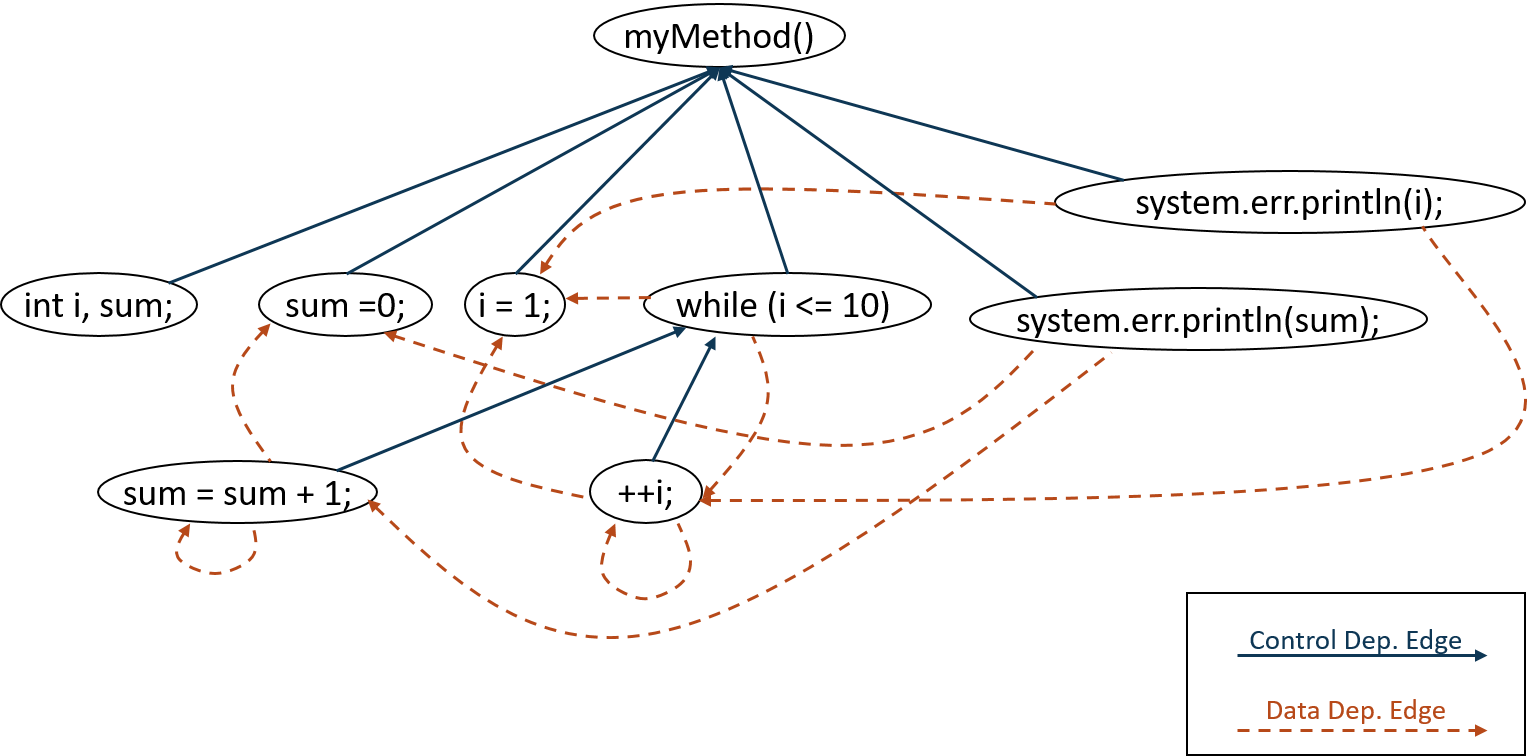}
		\caption{The transpose graph of the dependency call graph in Figure \ref{fig:dependency-call-graph example}}
		\label{fig:transpose-graph-example}
	\end{center}
\end{figure}

We rely of Algorithm \ref{algo:transpose-graph} to compute the transpose graph from the dependency call graph. Following this algorithm, Figure \ref{fig:transpose-graph-example} shows the transpose graph of the dependency call graph in Figure \ref{fig:dependency-call-graph example}.

\begin{algorithm}
\SetAlgoLined
\SetKwFunction{match}{match}
 \KwIn{Graph \textit{G = (V, E)}}
 \KwOut{Transpose Graph \textit{G$^{T}$ = (V, E$^{T}$)}}
  Graph \textit{G$^{T}$} = new Graph()\;
  \textit{G$^{T}$.V} = \textit{G.V}\;
  \For{each node \textit{u} $\in$ \textit{G.V}}
 {
 	\For{each \textit{e} $\in$ \textit{u.E}}
 	{
 		\textit{G$^{T}$.e}.append(\textit{u})\;
 	}
 }
\KwRet{\textit{G$^{T}$}\;}
\caption{Computing Transpose Graph}
\label{algo:transpose-graph}
\end{algorithm}

\subsection{Identifying Reachable Nodes Using Breadth-First Search Algorithm}
BFS allows us to identify all nodes that are reachable directly or indirectly from a given node.
We apply the Breadth-First Search (BFS) starting from the given node. Then, at each time, BFS will visit nodes at distance \textit{d} before nodes at distance \textit{d+1}. 

We propose Algorithm \ref{algo:bsf} that identifies a set of reachable nodes in a given graph \textit{G} and a given node \textit{n}. 

\begin{figure}[h]
	\begin{center}
		\includegraphics[width=\textwidth]{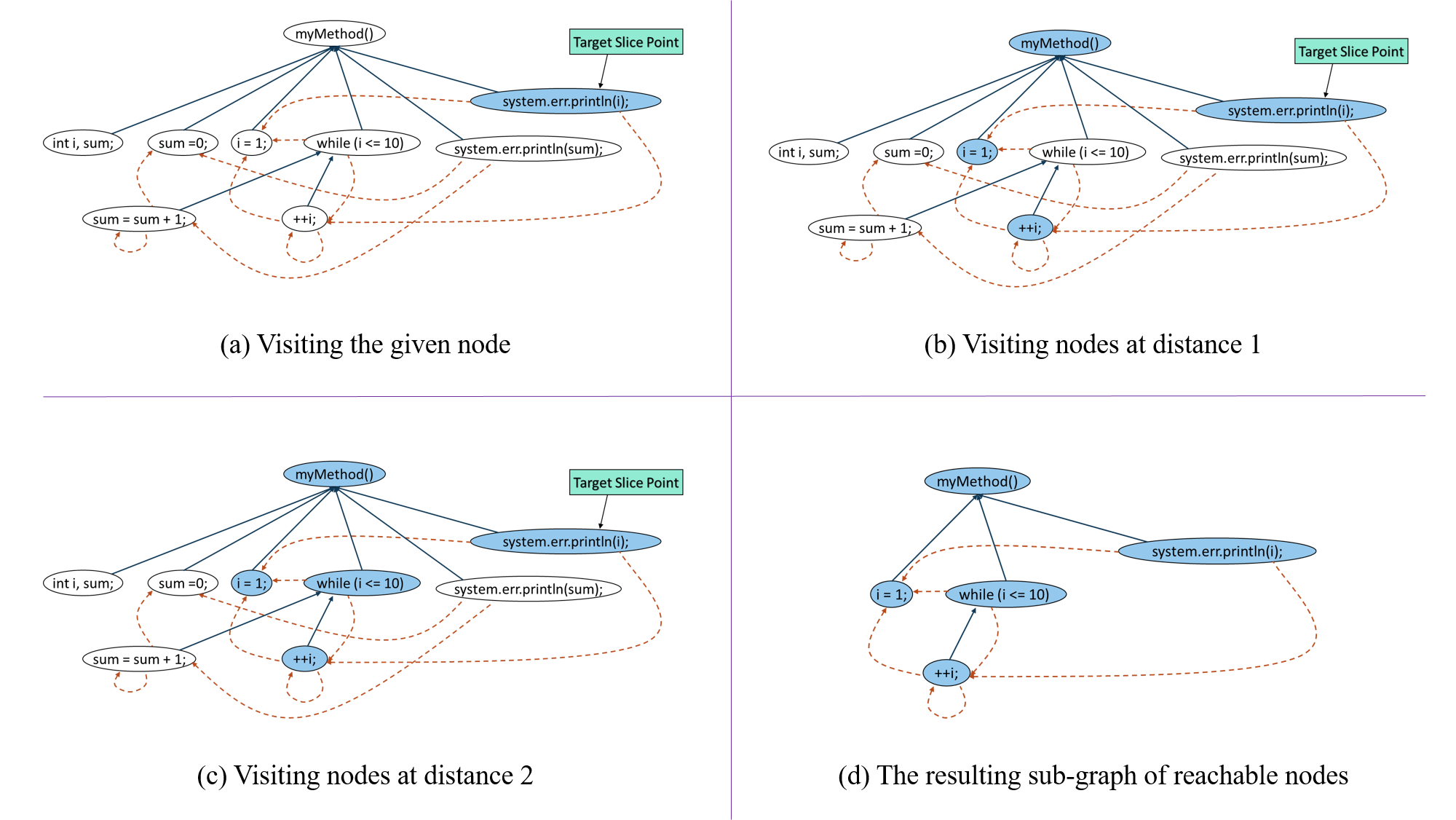}
		\caption{The use of BFS to identify reachable nodes in the transpose graph presented in Figure \ref{fig:transpose-graph-example}}
		\label{fig:bfs-example}
	\end{center}
\end{figure}

The procedure of this algorithm is illustrated in Figure \ref{fig:bfs-example} for the graph presented in Figure \ref{algo:transpose-graph} and the \textit{System.err.println(i);} node.

\begin{algorithm}
\SetAlgoLined
\SetKwFunction{match}{match}
 \KwIn{Graph \textit{G = (V, E)}, Target node \textit{n}}
 \KwOut{Set of reachable nodes \textit{nodes}}
  Set<Node> \textit{nodes} = new Set();
  Queue \textit{Q} = new Queue()\;
  \textit{Q}.enqueue(\textit{n})\;
  \textit{nodes}.append(\textit{n})\;
  \While{\textit{!Q}.isEmpty()}
  {
  	\textit{currentNode} = \textit{Q}.dequeue()\;
    \textit{adjacentNodes} = \textit{currentNode}.getFanOutNodes()\;
    \For{each node \textit{e} $\in$ \textit{adjacentNodes}}
 	{
    	\If{!\textit{nodes}.contiants(\textit{e})}
 			{
 				Q.enqueue(\textit{e})\;
				\textit{nodes}.append(\textit{e})\;
 			} 	
 	}
  }
\KwRet{\textit{nodes}\;}
\caption{BFS Algorithm to Identify Reachable Nodes}
\label{algo:bsf}
\end{algorithm}

\section{Tool Implementation}
\label{sec:tool-implementation}
We extend our DeJEE (Dependencies in JEE) tool proposed in \cite{shatnawi2017analyzing} to enable it to identify a program slice corresponding to an output stream object. 
We implement our approach based on the OMG's Knowledge Discovery Metamodel (KDM) representation \cite{perez2011knowledge} of the source code implementing a given  application. The idea behind that is to allow our approach to be language-independent one as it will not care about the underlying implementing programming languages of the given  application. We rely on the KDM APIs offered by the \textit{MoDisco}. We presented how to identify a KDM model for a given Java project based on the static analysis of its source code in our technical report in \cite{shatnawi2017:tch:kdm-from-jsp}. 

The complete Abstract Syntax Tree (AST) of all statements in the source code are included in the extracted KDM model. We develop an algorithm to identify a dependency call graph by parsing a given KDM model of a method. Then, we use this dependency call graph to identify the program slice in terms of KDM's \textit{ActionElement} instances that contain the real implementation. 

\section{Conclusion}
\label{sec:conclusion}
In this paper, we proposed a program slice approach that identifies a set of statements that may impact the value of the output stream object of Servlet/Tag Handler. Our approach identifies a dependency call graph based on the analysis of KDM models. Then, it identifies the program slice using the BFS algorithm as a set of reachable nodes in the transpose graph of the identified dependency call graph.

As future directions, we want to develop an approach that identifies a dependency call graph of a given KDM model based on rules that can be executed on top of this KDM model.

\bibliographystyle{unsrt}
\bibliography{main}  

\end{document}